# VANET Routing Protocols: Pros and Cons

### Bijan Paul
Dept. of Computer Science &
Engineering,
Shahjalal University of
Science & Technology,
Sylhet, Bangladesh

### Md. Ibrahim
Dept. of Computer Science &
Engineering,
Shahjalal University of
Science & Technology,
Sylhet, Bangladesh

### Md. Abu Naser Bikas
Lecturer, Dept. of Computer
Science & Engineering,
Shahjalal University of
Science & Technology,
Sylhet, Bangladesh

## ABSTRACT
VANET (Vehicular Ad-hoc Network) is a new technology which has taken enormous attention in the recent years. Due to rapid topology changing and frequent disconnection makes it difficult to design an efficient routing protocol for routing data among vehicles, called V2V or vehicle to vehicle communication and vehicle to road side infrastructure, called V2I. The existing routing protocols for VANET are not efficient to meet every traffic scenarios. Thus design of an efficient routing protocol has taken significant attention. So, it is very necessary to identify the pros and cons of routing protocols which can be used for further improvement or development of any new routing protocol. This paper presents the pros and cons of VANET routing protocols for inter vehicle communication.

## 1. INTRODUCTION
Vehicular ad hoc network is a special form of MANET which is a vehicle to vehicle & vehicle roadside wireless communication network. It is autonomous & self-organizing wireless communication network, where nodes in VANET involve themselves as servers and/or clients for exchanging & sharing information. The network architecture of VANET can be classified into three categories: pure cellular/WLAN, pure ad hoc, and hybrid [1]. Due to new technology it has taken huge attention from government, academy & industry. There are many research projects around the world which are related with VANET such as COMCAR [2], DRIVE [3], FleetNet [4] and NoW (Network on Wheels) [5], CarTALK 2000 [6], CarNet [7]. Figure-1 shows a form of vehicular adhoc network. There are several VANET applications such as Vehicle collision warning, Security distance warning, Driver assistance, Cooperative driving, Cooperative cruise control, Dissemination of road information, Internet access, Map location, Automatic parking, Driverless vehicles.

This paper summarizes the pros and cons of unicast routing protocols which can be used for better understanding of the routing protocols and future improvement can be made. The remainder of the paper is organized as follows: Section 2 describes the VANET characteristics. Section 3 discusses related research work on routing protocol design as applied to VANET. Section 4 & 5 presents the pros & cons of Topology based routing protocols & Position based routing protocols. We conclude in Section 6 and section 7 for reference.

## 2. CHARACTERISTICS
VANET has some unique characteristics which make it different from MANET as well as challenging for designing VANET applications.

### 2.1 High dynamic topology
The topology of VANET changes because of the movement of vehicles at high speed. Suppose two vehicles are moving at the speed of 20m/sec and the radio range between them is 160 m. Then the link between the two vehicles will last 160/20 = 8 sec.

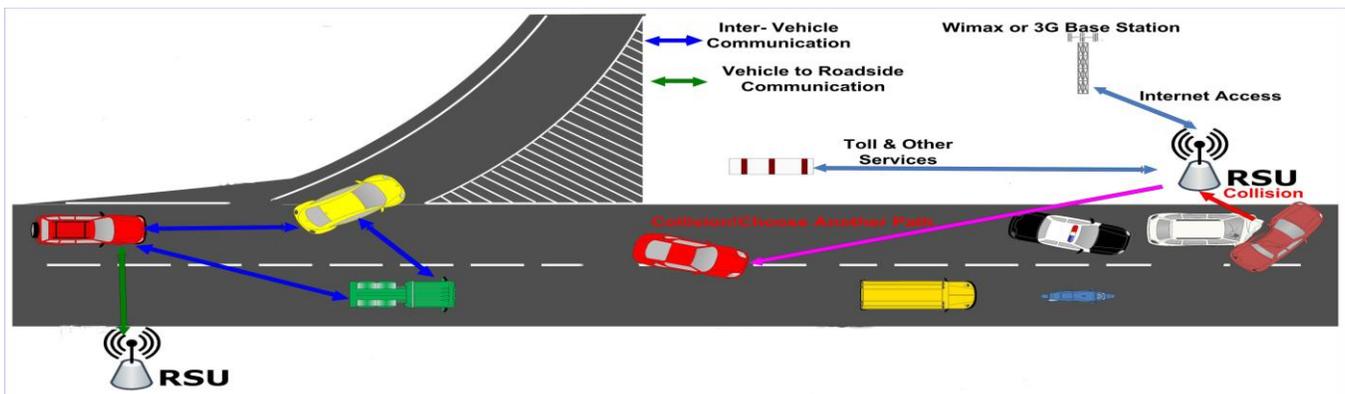

**Fig 1: Vehicular adhoc networks and some possible applications.**

## 2.2 Frequent disconnected network

From the highly dynamic topology results we observe that frequent disconnection occur between two vehicles when they are exchanging information. This disconnection will occur most in sparse network.

## 2.3 Mobility modeling

The mobility pattern of vehicles depends on traffic environment, roads structure, the speed of vehicles, driver's driving behavior and so on.

## 2.4 Battery power and storage capacity

In modern vehicles battery power and storage is unlimited. Thus it has enough computing power which is unavailable in MANET. It is helpful for effective communication & making routing decisions.

## 2.5 Communication environment

The communication environment between vehicles is different in sparse network & dense network. In dense network building, trees & other objects behave as obstacles and in sparse network like high-way this things are absent. So the routing approach of sparse & dense network will be different.

## 2.6 Interaction with onboard sensors

The current position & the movement of nodes can easily be sensed by onboard sensors like GPS device. It helps for effective communication & routing decisions.

## 3. ROUTING PROTOCOLS

The characteristic of highly dynamic topology makes the design of efficient routing protocols for VANET is challenging. The routing protocol of VANET can be classified into two categories such as Topology based routing protocols & Position based routing protocols. Overall classification of VANET routing protocols has been shown in the figure-2.

## 4. PROS & CONS OF TOPOLOGY BASED ROUTING PROTOCOLS

Topology based routing protocols use link's information within the network to send the data packets from source to destination. Topology based routing approach can be further categorized into proactive (table-driven) and reactive (on-demand) routing.

### 4.1 Proactive (table-driven)

Proactive routing protocols are mostly based on shortest path algorithms. They keep information of all connected nodes in form of tables because these protocols are table based. Furthermore, these tables are also shared with their neighbors. Whenever any change occurs in network topology, every node updates its routing table.

***Pros***

- No Route Discovery is required.
- Low Latency for real time applications.

***Cons***

- Unused paths occupy a significant part of the available bandwidth.

#### 4.1.1 Fisheye State Routing

FSR [8] is a proactive or table driven routing protocol where the information of every node collects from the neighboring nodes. Then calculate the routing table. It is based on the link state routing & an improvement of Global State Routing.

***Pros***

- FSR reduces significantly the consumed bandwidth as it exchanges partial routing update information with neighbors only.
- Reduce routing overhead.
- Changing in the routing table will not occur even if there is any link failure because it doesn't trigger any control message for link failure.

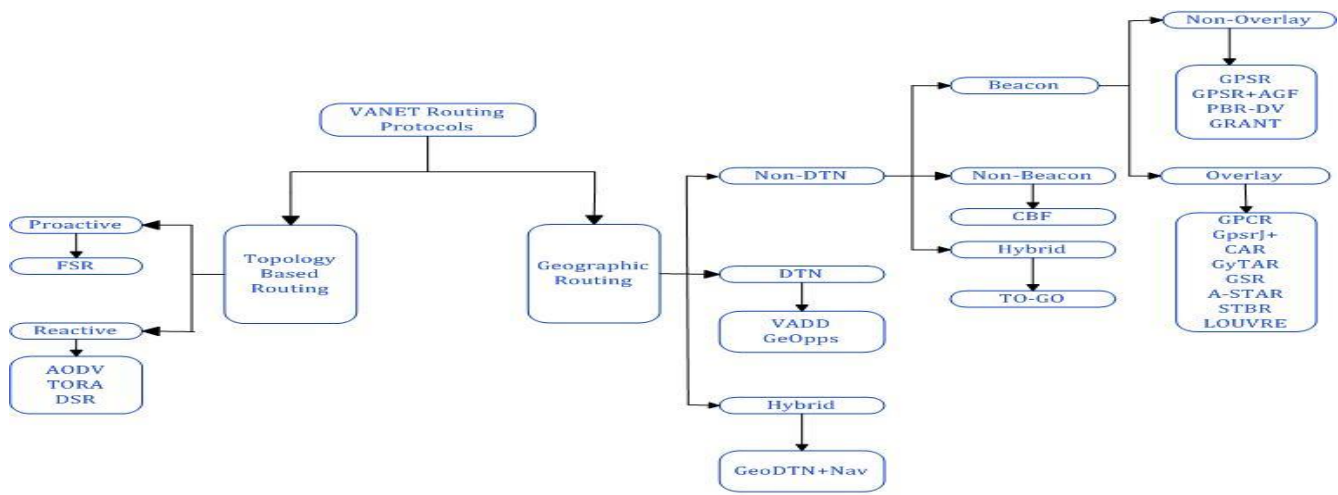

**Fig2: Unicast routing protocols in VANET**

***Cons***

-Very poor performance in small ad hoc networks.
-Less knowledge about distant nodes.
-The increase in network size the storage complexity and the processing overhead of routing table also increase.
- Insufficient information for route establishing.

## 4.2  Reactive (On Demand)

Reactive routing protocol is called on demand routing because it starts route discovery when a node needs to communicate with another node thus it reduces network traffic.

***Pros***

-To update routing table not require periodic flooding the network. Flooding requires when it is demanded.
-Beaconless so it saves the bandwidth.

***Cons***

- For route finding latency is high.
- Excessive flooding of the network causes disruption of nodes communication.

### 4.2.1  *AODV*

Ad Hoc On Demand Distance Vector routing protocol [9] is a reactive routing protocol which establish a route when a node requires to send data packets. It has the ability of unicast & multicast routing. It uses a destination sequence number (DestSeqNum) which makes it different from other on demand routing protocols.

***Pros***

- An up-to-date path to the destination because of using destination sequence number.
- It reduces excessive memory requirements and the route redundancy.
- AODV responses to the link failure in the network.
- It can be applied to large scale adhoc network.

***Cons***

-More time is needed for connection setup & initial communication to establish a route compared to other approaches.
-If intermediate nodes contain old entries it can lead inconsistency in the route.
-For a single route reply packet if there has multiple route reply packets this will lead to heavy control overhead.
- Because of periodic beaconing it consume extra bandwidth.

### 4.2.2  *Dynamic Source Routing*

The Dynamic Source Routing (DSR) protocol presented in [10] which utilize source routing & maintain active routes. It has two phases route discovery & route maintenance.

***Pros***

-Beacon less.

 -To obtain route between nodes, it has small overload on the network. It uses caching which reduce load on the network for future route discovery.
-No periodical update is required in DSR.

***Cons***

-If there are too many nodes in the network the route information within the header will lead to byte overhead.
-Unnecessary flooding burden the network.
-In high mobility pattern it performs worse.
-Unable to repair broken links locally.

### 4.2.2  *Temporally Ordered Routing Protocol (TORA)*

Temporally Ordered Routing Protocol [11] is based on the link reversal algorithm that creates a direct acyclic graph towards the destination where source node acts as a root of the tree. In TORA packet is broadcasted by sending node, by receiving the packet neighbor nodes rebroadcast the packet based on the DAG if it is the sending node's downward link.

***Pros***

-It creates DAG (Direct acyclic graph) when necessary.
-Reduce network overhead because all intermediate nodes don't need to rebroadcast the message.
-Perform well in dense network.

***Cons***

-It is not used because DSR & AODV perform well than TORA.
-It is not scalable.

## 5. PROS & CONS OF GEOGRAPHIC ROUTING PROTOCOLS

Geographic routing is a routing that each node knows it's own & neighbor node geographic position by position determining services like GPS. It doesn't maintain any routing table or exchange any link state information with neighbor nodes. Information from GPS device is used for routing decision.

***Pros***

- Route discovery & management is not required.
-Scalability.
-Suitable for high node mobility pattern.

***Cons***

-It requires position determining services.
-GPS device doesn't work in tunnel because satellite signal is absent there.

We need to know some terms that we have used in the figure-2 & then we will describe the pros & cons of geographic routing protocols as shown in the figure-2.

## 5.1  DTN

Delay Tolerant Network (DTN) uses carry & forward strategy to overcome frequent disconnection of nodes in the network. In carry & forward strategy when a node can't contact with other nodes it stores the packet & forwarding is done based on some metric of nodes neighbors.

## 5.2 BEACON

Beacon means transmitting short hello message periodically. It exposes presence and position of a node. An entry will be removed from neighbor table of a receiving node if it fails to receive a beacon after a certain period of time from the corresponding node.

## 5.3         OVERLAY

Overlay is a network that every node is connected by virtual or logical links which is built on top of an existing network.

### 5.1.1 VADD (Vehicle-Assisted Data Delivery)

Vehicle-Assisted Data Delivery [12] is based on the idea of carry & forward approach by using predicable vehicle mobility. Among proposed VAAD protocols H-VAAD shows better performance.

**Pros**

-Comparing with GPSR (with buffer), epidemic routing and DSR, VADD performs high delivery ratio.
-It is suitable for multi-hop data delivery.

**Cons**

- Due to change of topology & traffic density it causes large delay.

### 5.1.2 Geographical Opportunistic Routing (GeOpps)

Geographical Opportunistic Routing (GeOpps) [13] protocol utilizes the navigation system suggested routes of vehicles for selecting the forwarding node which is closer to the destination. During this process if there is any node which has minimum arrival time the packet will be forwarded to that node.

**Pros**

-By comparing with the Location-Based Greedy routing and MoVe routing algorithm GeOpps has high delivery ratio.
-To find a vehicle which is driving towards near the destination GeOpps need few encounters.
- The delivery ratio of GeOpps rely on the mobility patterns & the road topology but not dependent on high density of vehicles.

**Cons**

-Privacy is an issue because navigation information is disclosed to the network.

### 5.2.1 Greedy Perimeter Stateless Routing (GPSR)

Greedy Perimeter Stateless Routing [14] selects a node which is closest to the final destination by using beacon. It uses greedy forwarding algorithm if it fails it uses perimeter forwarding for selecting a node through which a packet will travel.

**Pros**

-To forward the packet a node needs to remember only one hop neighbor location.
-Forwarding packet decisions are made dynamically.

**Cons**

-For high mobility characteristics of node, stale information of neighbors' position are often contained in the sending nodes' neighbor table.
-Though the destination node is moving its information in the packet header of intermediate node is never updated.

### 5.2.2 GPSR+AGF

In GPSR we see that stale information of neighbors position are often contained in the sending nodes neighbor table. For this reason an approach which is called Advanced Greedy Forwarding (AGF) [15] is proposed.

**Pros**

- Though the destination node is moving its information in the packet header of intermediate node is updated.
-Stale nodes of neighbor table can be detected.

**Cons**

- To find the shortest connected path it may not give desired optimal solution.

### 5.2.3 PBR-DV

PBR-DV consists of various approaches such as a greedy, position-based and a reactive, topology-based routing strategy & if packet falls in local maximum it uses AODV approach recovery.

**Pros**

- No comparison is done with any other routing protocol so uncertain about packet delivery ratio & overhead.

**Cons**

-For non-greedy part excessive flooding is required.

### 5.2.4 Greedy Routing with Abstract Neighbor Table (GRANT)

To avoid local maximum Greedy Routing with Abstract Neighbor Table (GRANT) [16] applies extended greedy routing algorithm concept. Abstract Neighbor Table of GRANT divides the plane into areas and includes per area only one representative neighbor.

**Pros**

-In city scenario with obstacles this extended greedy routing approach works well than as usual greedy approach.

**Cons**

- VANET has a high mobility characteristics but the performance evaluation of GRANT is done on static traces.
-The overhead of beacon and possible inaccuracy in packet delivery are not measured.

### 5.3.1 Greedy Perimeter Coordinator Routing (GPCR)

Greedy Perimeter Coordinator Routing [17] is a position-based routing protocol uses greedy algorithms to forward packet based on a pre-selected path which has been designed to deal with the challenges of city scenarios. No global or external information like static map does not require in GPCR.

### Pros

- Does not require any global or external information.
-For representing the planar graph it uses the underlying roads though it is based on the GPSR.
-It has no as usual a planarization problem like unidirectional links, planar sub-graphs & so on.

### Cons

- Depends on junction nodes.
-There has a problem in the Junction detection approach in which first approach fails on curve road & second approach fails on a sparse road.

### 5.3.2                                    GpsrJ+

GpsrJ+ [18] is a position based routing protocol which reduces the dependency on junction node. By using digital maps GpsrJ+ recovers from the local maximum. It uses two hop neighbors information for detecting appropriate junction turns & to calculate a good routing path.

### Pros

-The packet delivery ratio of GPCR increases which is managed by GPSRJ+.
-The number of hops in the recovery mode of GPSR is reduced by 200%.
- An expensive planarization strategy is not required in GPSRJ+.

### Cons

- Not appropriate for the delay sensitive applications.
- It did not apply on realistic city map that are not necessarily grids.
- It has used simple line trajectory but realistic roads follow a more complex trajectory.

### 5.3.3    CAR    (Connectivity-Aware    Routing)

For city and/or highway environment Connectivity-Aware Routing (CAR) [19] is designed which uses AODV for path discovery and uses PGB for data dissemination mode. It uses guard concept to maintain the path.

### Pros

-No digital map is required.
-It has no local maximum problem.
- CAR ensures to find the shortest connected path because CAR has higher packet delivery ratio than GPSR and GPSR+AGF.

### Cons

-Unnecessary nodes can be selected as an anchor.
-It cannot adjust with different sub-path when traffic environment changes.

### 5.3.4 Greedy Traffic Aware Routing protocol (GyTAR)

Greedy Traffic Aware Routing protocol [23] gives a new concept of intersection-based routing protocol which aims to reduce the control message overhead & end-to-end delay with low packet loss.

### Pros

- For high mobility topology changes rapidly and often occurring network fragmentation which is efficiently handle by GyTAR.
-Performance shows that throughput, delay and routing overhead are better than GSR.

### Cons

- GyTAR depends on roadside units because it assumes that the number of cars in the road will be given from road side units.
-Gytar cannot avoid void.

### 5.3.5    GSR(Geographic    Source    Routing)

GSR [21] routing was proposed for vehicular ad hoc networks in city environments which is the combination of position-based routing with topological knowledge.GSR uses greedy forwarding along a pre-selected shortest path & this path is calculated by using Dijkstra algorithm.

### Pros

- Packet delivery ratio of GSR is better than AODV & DSR.
- GSR is scalable than AODV & DSR.

### Cons

-This protocol neglects the situation like sparse network where there are not enough nodes for forwarding packets.
-GSR shows higher routing overhead GyTAR because of using hello messages as control messages.

### 5.3.6 Anchor-Based Street and Traffic Aware Routing (A-STAR)

Anchor-Based Street and Traffic Aware Routing [22] (A-STAR) is a position based routing protocol which is specially design for city scenarios for inter vehicle communication system. It ensures high connectivity in packet delivery by using vehicular traffic city bus information for an end-to-end connection.

### Pros

-In low traffic density, A-STAR ensures for finding an end-to-end connection.

-By comparing with the greedy approach of GSR & the perimeter mode of GPSR. A-STAR uses a new local recovery strategy which is more suitable for city environment.
-Path selection of A-STAR ensures high connectivity though its packet delivery ratio is lower than GSR & GPSR.

### Cons

-Packet delivery ratio of A-STAR is lower than GSR & GPSR.
- To find a path from source to destination it uses static information based on city bus routes which causes connectivity problem on some portion of streets.

### 5.3.7 Street Topology Based Routing (STBR)

Street Topology-Based Routing(STBR) [23] is based on the idea

of elucidate a given street map as a planar graph which has three valid states: master, slave, and forwarder for a node. In STBR one node is selected as a master on a junction, other nodes act as slaves & intermediate nodes between junctions act as forwarders.

### Pros

-It traverses least spanning multiple junctions for long distance unicast communication.

### Cons

- STBR is not appropriate for mixed scenarios because it would try to send junction beacons along a highway.
- In STBR complexity increases because of some special cases like transferring the two-hop neighbor table to the new master when the old master leaves the junction .

### 5.3.8 LOUVRE (Landmark Overlays for Urban Vehicular Routing Environments)

LOUVRE [24] is a geo-proactive overlay routing protocol which ensures an obstacle-free routing on the overlay links and also reduces the chances of falling into a local maximum.

### Pros

- Avoid void and backtracking because of using estimation of Peer-to-peer density.
- Packet delivery ratio is higher than GPCR & GPSR.
- Ensures an obstacle free geographic routing.

### Cons

-Because of delivering unsuccessful packets it has a little higher hop count than GPCR.
 -Scalability.

### CBF (Contention-Based Forwarding)

Contention-Based Forwarding [25] is a geographic routing protocol that does not make use of beacons. In CBF if there is a data packet to send, the sending node will broadcast the packet to all direct neighbors & these neighbors will find out among themselves the one that will forward the packet.

### Pros

-Elimination of beacon message saves bandwidth.
-Reduces the probability of packet collision & inefficient routing by ignoring inaccurate neighbor tables.
-When node mobility is high CBF protocol provides a lower packet forwarding delay.

### Cons

- In high way destination is always straight forward so local maximum never occurs as a result CBF works well but in city environment local maximum frequently occurs because source and destination may lie on different path.

### TO-GO(Topology-assist Geo-Opportunistic Routing)

TO-GO [26]  is a geographic routing protocol which improves packet delivery in greedy & recovery forwarding that can bypass the junction area by using two hop beaconing.

### Pros

-No hidden terminal occurs because all nodes can hear one another.
-From simulation result TO-GO, GPCR, GpsrJ+  have similar packet delivery ratio.
- Low S/N ratio is taken care of.

### Cons

-Simulation result shows that End-to-End latency in TO-GO is higher than GPCR, GPSR, GpsrJ+.

### GeoDTN+Nav

GeoDTN+Nav [27] is a combination of DTN & Non-DTN mode which includes a greedy mode, a perimeter mode and a DTN mode. It can switch from Non-DTN to DTN mode. This approach proposes virtual navigation interface (VNI) which provides necessary information for GeoDTN+Nav to determine its routing mode and forwarder.

### Pros

-GeoDTN+Nav can switch from Non-DTN to DTN mode.
-GeoDTN+Nav can recognize partition in the network.

### Cons

-The latency increases & the decreases packet delivery ratio in a situation such as sparse network where GeoDTN+Nav tries to fall-back to DTN mode again.
-The result in a partitioned network shows that RandDTN achieves slightly better PDR and lower latency than GeoDTN+Nav.

## 6. CONCLUSION

In this paper, we have investigated the pros and cons of different routing protocols for inter-vehicle communication in VANET. By studying different routing protocol in VANET we have seen that further performance evaluation is required to verify performance of a routing protocol with other routing protocols based on various traffic scenarios. Comparison can be done among the routing protocols in the Overlay and so on. GSR is not compared with other position based routing protocol. Besides, performance evaluation of PBR-DV is not done with the non-overlay routing protocols.

## 7. REFERENCES

[1]  "Survey of Routing Protocols in Vehicular Ad Hoc Networks," Kevin C. Lee, Uichin Lee, Mario Gerla, Advances in Vehicular Ad-Hoc Networks: Developments and Challenges, IGI Global, Oct, 2009.

[2]  Ericson, "Communication and Mobility by Cellular Advanced Radio", ComCar project, www.comcar.de, 2002.

[3]  Online, http://www.ist-drive.org/index2.html.

[4]   W. Franz, H. Hartenstein, and M. Mauve, Eds., *Inter-Vehicle-Communications Based on Ad Hoc Networking Principles-The Fleet Net Project.* Karlshue, Germany: Universitatverlag Karlsuhe,November 2005.

[5]   A. Festag, et. al., "NoW-Network on Wheels: Project Objectives,Technology and Achievements", Proceedings

of 6th InternationalWorkshop on Intelligent Transportations (WIT), Hamburg, Germany,March 2008.

[6] Reichardt D., Miglietta M., Moretti L., Morsink P., and Schulz W.,"CarTALK 2000 − safe and comfortable driving based upon inter-vehicle-communication," in Proc. IEEE IV'02.

[7] Morris R., Jannotti J., Kaashoek F., Li J., Decouto D., "CarNet: A scalable ad hoc wireless network system," 9th ACM SIGOPS European Workshop, Kolding, Denmark, Sept. 2000.

[8] Pei, G., Gerla, M., and Chen, T.-W. (2000), "Fisheye State Routing: A Routing Scheme for Ad Hoc Wireless Networks," Proc. ICC 2000, New Orleans, LA, June 2000.

[9] Perkins, C.; Belding-Royer, E.; Das, S. (July 2003)"Ad hoc On-Demand Distance Vector (AODV) Routing".

[10] Johnson, D. B. and Maltz, D. A. (1996), "Dynamic Source Routing in Ad Hoc Wireless Networks," Mobile Computing, T. Imielinski and H. Korth, Eds., Ch. 5, Kluwer, 1996, pp. 153–81.

[11] Park, V.D., Corson, M.S. (1997), "A highly adaptive distributed routing algorithm for mobile wireless networks," INFOCOM '97. Sixteenth Annual Joint Conference of the IEEE Computer and Communications Societies. Proceedings IEEE , vol.3, no., pp.1405-1413 vol.3, 7-12 Apr 1997.

[12] Zhao, J.; Cao, G. (2006), "VADD: Vehicle-Assisted Data Delivery in Vehicular Ad Hoc Networks," INFOCOM 2006. 25th IEEE International Conference on Computer Communications. Proceedings , vol., no., pp.1-12, April 2006.

[13] Leontiadis, I., Mascolo, C. (2007), "GeOpps: Geographical Opportunistic Routing for Vehicular Networks," World of Wireless, Mobile and Multimedia Networks, 2007. WoWMoM 2007. IEEE International Symposium on a , vol., no., pp.1-6, 18-21 June 2007.

[14] Karp, B. and Kung, H. T (2000), "GPSR: greedy perimeter stateless routing for wireless networks." In Mobile Computing and Networking, pages 243-254, 2000.

[15] Naumov, V., Baumann, R., Gross, T. (2006), "An evaluation of Inter-Vehicle Ad Hoc Networks Based on Realistic Vehicular Traces," Proc. ACM MobiHoc'06 Conf., May, 2006.

[16] Schnaufer, S., Effelsberg, W. (2008), "Position-based unicast routing for city scenarios," World of Wireless, Mobile and Multimedia Networks, 2008. WoWMoM 2008. 2008 International Symposium on a , vol., no., pp.1-8, 23-26 June 2008.

[17] Lochert, C., Mauve, M., F`ussler, H., and Hartenstein, H., "Geographic routing in city scenarios," SIGMOBILE Mob. Comput. Commun. Rev., vol. 9, no. 1, pp. 69–72, 2005.

[18] Lee, K. C., Haerri, J., Lee, U., and Gerla, M. (2007), "Enhanced perimeter routing for geographic forwarding protocols in urban vehicular scenarios," Globecom Workshops, 2007 IEEE, pp. 1–10, 26-30 Nov. 2007.

[19] Naumov, V., Gross, T.R. (2007), "Connectivity-Aware Routing (CAR) in Vehicular Ad-hoc Networks," INFOCOM 2007. 26th IEEE International Conference on Computer Communications. IEEE , vol., no., pp.1919-1927, 6-12 May, 2007.

[20] Jerbi, M., Senouci, S.-M., Meraihi, R., and Ghamri-Doudane, Y. (2007), "An improved vehicular ad hoc routing protocol for city environments," Communications, 2007. ICC '07. IEEE International Conference, pp. 3972–3979, 24-28 June 2007.

[21] Lochert, C., Hartenstein, H., Tian, J., Fussler, H., Hermann, D., Mauve, M. (2003), "A routing strategy for vehicular ad hoc networks in city environments," Intelligent Vehicles Symposium, 2003. Proceedings. IEEE , vol., no., pp. 156-161, 9-11 June 2003.

[22] Seet, B.-C., Liu, G., Lee, B.-S., Foh, C. H., Wong, K. J., Lee, K.-K. (2004), "A-STAR: A Mobile Ad Hoc Routing Strategy for Metropolis Vehicular Communications." NETWORKING 2004, 989-999.

[23] Forderer, D (2005). "Street-Topology Based Routing." Master's thesis, University of Mannheim, May 2005.

[24] Lee, K., Le, M., Haerri J., and Gerla, M. (2008), "Louvre: Landmark overlays for urban vehicular routing environments," Proceedings of IEEE WiVeC, 2008.

[25] F`ußler, H., Hannes, H., J¨org, W., Martin, M., Wolfgang, E. (2004), "Contention-Based Forwarding for Street Scenarios," Proceedings of the 1st International Workshop in Intelligent Transportation (WIT 2004), pages 155–160, Hamburg, Germany, March 2004.

[26] Lee, K.C.; Lee, U.; Gerla, M. (2009), "TO-GO: TOpology-assist geo-opportunistic routing in urban vehicular grids," Wireless On-Demand Network Systems and Services, 2009. WONS 2009. Sixth International Conference on , vol., no., pp.11-18, 2-4 Feb. 2009.

[27] Cheng, P.-C., Weng, J.-T., Tung, L.-C., Lee, K. C., Gerla M., and Härri J. (2008), "GeoDTN+NAV: A Hybrid Geographic and DTN Routing with Navigation Assistance in Urban Vehicular Networks," Proceedings of the 1st International Symposium on Vehicular Computing Systems (ISVCS'08), Dublin, Irland, July.